\documentclass[12pt]{article}
\usepackage{amssymb}
\usepackage{epsfig}

\newcommand{\sweffl}{\sin^2\theta_{W,eff}}
\newcommand{\sweff}{s^2_W}
\newcommand{\sweffq}{s^4_W}
\newcommand{\sweffu}{s_W}

\newcommand{\bq}{\begin{equation}}
\newcommand{\eq}{\end{equation}}
\newcommand{\bqa}{\begin{eqnarray}}
\newcommand{\eqa}{\end{eqnarray}}
\newcommand{\ben}{\begin{enumerate}}
\newcommand{\een}{\end{enumerate}}
\newcommand{\bc}{\begin{center}}
\newcommand{\ec}{\end{center}}

\def\pr#1#2#3{ Phys. Rev. ${\bf{#1}}$ (#2) #3}

\def\pl#1#2#3{ Phys. Lett. ${\bf{#1}}$ (#2) #3}

\def\np#1#2#3{ Nucl. Phys. ${\bf{#1}}$ (#2) #3}
\def\zp#1#2#3{ Z. f. Phys. ${\bf{#1}}$ (#2) #3}

\def\etal{{\it et.al.\/}}

\global\nulldelimiterspace = 0pt

\def\L{ {\cal L }}
\def\O{ {\cal O }}

\begin{document}
\thispagestyle{empty}
\begin {flushleft}
 PM/98-19\\
RAL-TR-1998-068 \\

October 1998\\
\end{flushleft}

\vspace*{2cm}

\hspace*{-0.5cm}
\begin{center}
{\Large {\bf  Bounds on anomalous gauge couplings}}\\ 
{\Large {\bf from past and
near future experiments:}}\\
{\Large {\bf ~~~~~~~~~the role of the different measurements}}
\footnote{Partially 
supported by the EC contract CHRX-CT94-0579.}\hspace{2.2cm}\null 
\hspace*{-0.5cm}
\vspace{1.cm} \\{\large M. Beccaria$^a$, F.M.
Renard$^b$,
S. Spagnolo$^c$ and C. Verzegnassi$^a$}\hspace{2.2cm}\null
\vspace {0.5cm} \\
\begin{center}
$^a$Dipartimento di Fisica, Universit\`a di 
Lecce and INFN, Sezione di Lecce\\
Via Arnesano, 73100 Lecce, Italy.\\
\vspace{0.2cm} 
$^b$ Physique
Math\'{e}matique et Th\'{e}orique, UMR 5825\\
Universit\'{e} Montpellier
II,  F-34095 Montpellier Cedex 5.\hspace{2.2cm}\\
\vspace{0.2cm} 
$^c$ 
Rutherford Appleton Laboratory - Particle Physics Department \\
Chilton, Didcot, Oxfordshire OX11 0QX
\end{center}

\vspace*{3cm}
{\bf Abstract}\hspace{2.2cm}\null
\end{center}
\hspace*{-1.2cm}
\begin{minipage}[b]{16cm}
We analyze the bounds on the set of four "non-blind" anomalous gauge
couplings that will be derived, in absence of deviations from the
Standard Model predictions, from near future measurements at LEP2 and
at the TEVATRON. The bounds are obtained by combining these negative
results with those already available from LEP1 and from atomic
parity violation experiments. In this process, the information coming
from LEP2 is treated 
using a recently proposed ("$Z$-peak subtracted") theoretical 
approach. This makes
it easier to identify the specific role that the different
experiments play in this determination, in the spirit of a recent
previous investigation for the set of three "blind" couplings.
\end{minipage} 

\setcounter{footnote}{0} 
\clearpage
\newpage 
  
\hoffset=-1.46truecm
\voffset=-2.8truecm
\textwidth 16cm
\textheight 22cm
\setlength{\topmargin}{1.5cm}

\section{Introduction.} 

The possibility that anomalous gauge couplings (AGC) exist, that
essentially respect the same gauge symmetry of the Standard Model, has
been considered in recent years by several authors \cite{anom}. In
particular, a classification scheme that differentiates the so-called
"blind" operators from the "non-blind" ones has been proposed
\cite{DeR}. This was reformulated in a general form by Hagiwara et al
in a work \cite{Hag} whose notations we shall follow in the
present paper.\par 

In a previous publication \cite{gm2}, we have analyzed the overall
bounds that can be obtained on the "blind" set by a combination of
(supposedly negative) experimental results, obtained in the two $W$
channel at LEP2 \cite{LEP2}, with those available from the LEP1
measurements of the $Z$ partial width into $b\bar b$ pairs \cite{LEP1}
and from the hypothetical future improved measurement of the muon
$g-2$ at BNL \cite{BNL}. A nice feature of that analysis, in our
opinion, was the fact that it pointed out in a clear way the fact that
the three different experiments were complementary, each one allowing
an improved determination of a different subset of parameters.\par
The aim of this short paper is that of showing that an essentially
similar situation can be obtained for the case of the "non-blind" set.
Here the available experimental information will be derived from
LEP1 measurements, from Atomic Parity Violation (APV)
results, from an assumed high precision measurement of the $W$ mass
at LEP2 and at the TEVATRON
and from three measurements that are being performed at
LEP2 in the final two fermion channel (those of the final muons cross
section and forward-backward asymmetry and that of the final hadronic
states cross section). For what concerns the latter information from
LEP2, we shall use a theoretical description based on the so-called
"$Z$-peak subtracted" approach \cite{Zsub}. This will allow to treat
the AGC contribution in a particularly simple way, thus making the
"genuine" role of the separate experiments in this procedure relatively
simple to identify.\par
We now proceed to illustrate our approach. The relevant "non-blind" part
of the anomalous Lagrangian can be written in the following way 
\cite{Hag} (only
dimension six operators are retained):

\bq
\L^{(NB)}= {f_{DW}\over\Lambda^2}\O_{DW}+{f_{DB}\over\Lambda^2}\O_{DB}
+{f_{BW}\over\Lambda^2}\O_{BW}+{f_{\Phi,1}\over\Lambda^2}\O_{\Phi,1}
\eq
\bqa
\O_{DW} & =& Tr ([D_{\mu},{\overrightarrow{W}}_{\nu
\rho})] [D^{\mu},{\overrightarrow{W}}^{\nu \rho}])  \ \ \
  , \ \  \label{listDW}  \\[0.1cm]
\O_{DB} & = & -{g'^2\over2}(\partial_{\mu}B_{\nu \rho})
(\partial^\mu B^{\nu
\rho}) \ \ \  , \ \   \label{listDB} \\[0.1cm] 
\O_{BW} & =&  \Phi^\dagger {B}_{\mu \nu}
\overrightarrow \tau \cdot {\overrightarrow{W}}^{\mu \nu} \Phi
\ \ \  , \ \  \label{listBW}  \\[0.1cm] 
\O_{\Phi 1} & =& (D_\mu \Phi^\dagger \Phi)( \Phi^\dagger
D^\mu \Phi) \ \ \  , \ \       \label{listPhi1}   
\eqa

The contribution of this Lagrangian to the LEP1 observables can be
easily computed at the tree level. At the one loop level in wich we
shall be interested, it has been shown in Ref.\cite{Hag} 
that for massless
fermions all the relevant AGC effects can be simply written by formally
replacing the four parameters that enter eq.(1) by corresponding
"renormalized" quantities. These are four combinations of each one of
the parameters of eq.(1) with contributions coming from the "blind"
set, and their expressions can be found in Ref.\cite{Hag}. In terms of
these four renormalized "non-blind" parameters, denoted as $f^r_{DW}$,
$f^r_{DB}$, $f^r_{BW}$, $f^r_{\Phi,1}$, the expressions of the various
observables at one loop can be easily derived.\par
The previous treatment has one important situation  where it does not
apply. In the theoretical expression of the partial $Z$ width in $b\bar
b$ pairs, $\Gamma_b$, the dominant $\simeq m^2_t$ contribution (where
the top mass cannot be ignored) comes from parameters of the "blind"
set, as exhaustively discussed in a previous paper \cite{bbAGC}. This
fact was actually exploited in Ref.\cite{gm2} to compute bounds on the
corresponding parameter sector. For the aims of this paper, that are
orthogonal to those of Ref.\cite{gm2}, the experimental value of
$\Gamma_b$ will consequently not be exploitable. For the identical
reason, the value of the full $Z$ hadronic width $\Gamma_h$ will also
not be considered.\par
In practice, therefore, we shall be entitled to use the values of the
two independent purely leptonic observables, the $Z$ partial width into
(charged) leptons $\Gamma_l$ and the leptonic forward-backward
asymmetry $A_{FB,l}$. The contributions to these quantities from
anomalous gauge couplings have the following theoretical expressions:

\bqa
\frac{\Gamma^{AGC}_l}{M_Z} &=& \frac{M_Z^2}{\Lambda^2}\left[
-\frac{\sweff(1-\sweff)}{2\pi\alpha}\times \right. \nonumber \\
&\times& \left(
1+\frac{8\sweff(1-\sweff)(1-4\sweff)}{1-2\sweff}
\right)f^r_{\Phi, 1} +  \nonumber \\
&+& \frac{8\sweff(1-\sweff)}{1-2\sweff} f^r_{BW} + \\ 
&+& \frac{8\pi\alpha(1-\sweff)}{\sweff}\left(
\frac{8\sweffq(1-4\sweff)}{1-2\sweff}-1
\right) f^r_{DW} + \nonumber \\
&+& \left . \frac{8\pi\alpha\sweff}{1-\sweff}\left(
\frac{8(1-\sweff)^2(1-4\sweff)}{1-2\sweff}-1
\right) f^r_{DB}
\right]\nonumber
\eqa


\bqa
A^{AGC}_{FB,l}&=&{24(1-4\sweff)(1-(1-4\sweff)^2)\sweff\over
(1+(1-4\sweff)^2) (1-2\sweff)}\frac{M^2_Z}{\Lambda^2} \nonumber\\
&&\left[{(1-\sweff)^2 \sweff\over2\pi\alpha}f^r_{\Phi,1}+\right.\\
&+& \left. (1-\sweff) f^r_{BW}+8\pi\alpha
(1-\sweff)(f^r_{DW}+f^r_{DB})\right] \nonumber 
\eqa

\noindent
where $\sweff=\sweffl$ is the ``effective'' weak mixing angle in
the commonly used (LEP1, SLC) definition.

Eqs.(6,7) will be the LEP1 content of our analysis. The next
experimental result that we shall use is that coming from the
measurement of Atomic Parity Violation \cite{APV}. This is usually
expressed in terms of the "weak charge" $Q_W$. 
The contribution to this quantity from anomalous gauge couplings is:

\bq
Q^{AGC}_W=0.80\ {M^2_Z\over\Lambda^2}\ \frac{4\sweff (1-\sweff)}{\alpha}\ f^r_{BW}
\eq


Next, we have added the experimental information coming from future
measurements of the $W$ mass, for which the AGC effect reads:

\bqa
M^{AGC}_W&&=[16.18 GeV][{8\pi\alpha M^2_Z\over\Lambda^2}({{1-\sweff}\over
\sweff}f^r_{DW}+f^r_{DB})\nonumber\\
&&+{(1-\sweff)(1-4\sweff)^2\over2 
\sweff \Lambda^2}f^r_{\Phi,1}-
{2M^2_W\over\Lambda^2}f^r_{BW}]
\eqa

The final experimental input is that derived from the final two fermion
channel at LEP2. We shall use three measurements, i.e. those of the
muon cross section $\sigma_{\mu}$ and forward-backward asymmetry
$A_{FB,\mu}$ and that of the full final hadronic cross section
$\sigma_5$. This last choice might seem in contradiction with our
previous remark concerning $\Gamma_b$, $\Gamma_h$ but in fact it is not
so. The reason is that the theoretical expressions that we shall use
have been derived in the so-called "$Z$-peak subtracted" approach,
exhaustively illustrated in Ref.\cite{Zsub}. This consists,
essentially, of replacing the input parameter $G_{\mu}$ by quantities
directly measured on top of $Z$ resonance. As an immediate byproduct,
the theoretical expressions of those one loop corrections that modify
the simple $Z$ propagator become subtracted at $q^2=(p_{e^-}+p_{e^+})^2
=M^2_Z$ and loose those terms that are energy independent. This
happens, in particular, to the dominant part of the 
$\simeq m^2_t$ "blind" contribution
carried from $\sigma_b$ to $\sigma_5$ at LEP2 which is, so to say,
"reabsorbed" by the input parameters $\Gamma_5$, and for a full and
detailed discussion of this point we defer to Ref.\cite{Zsub}.\par
The elimination  of the "blind" parameters in $\sigma_5$ is not the
only bonus of the use of the "$Z$-peak subtracted" approach. For the
same reason that we have just mentioned, those "non blind"
renormalized parameters whose relevant contribution would be energy
independent \underline{disappear} in the subtracted expressions (this is also valid
for the contributions to the photon propagator, that are by definition
subtracted at $q^2=0$). As one can guess, this
applies to the two parameters $f^r_{BW}$, $f^r_{\Phi,1}$: these are
multiplied by the smaller number of derivatives in the associated
operators and consequently are reabsorbed, in our approach, 
in the LEP1 experimental measurements
used as new theoretical inputs. 
In conclusion, the AGC ``genuine LEP2'' contribution,
that is, we repeat, \underline{the one that cannot be
reabsorbed by the use of the LEP1 measurements}, reads:

\bqa  \sigma_{\mu}(q^2)&=&\sigma^{Born}_{\mu}(q^2)\bigm\{1+\nonumber\\
&&{2\over
\kappa^2(q^2-M^2_Z)^2+q^4}\left[\kappa^2(q^2-M^2_Z)^2
\tilde{\Delta}_{\alpha}(q^2)
-q^4(R(q^2)+0.5V(q^2))\right]\bigm\} \eqa
\noindent
where $\kappa\equiv{\alpha M_Z\over3\Gamma_l}\simeq2.64$ and
\bq
\sigma^{Born}_{\mu}(q^2)= {4\pi\alpha^2\over3q^2}
\left[{ q^4+\kappa^2(q^2-M^2_Z)^2\over\kappa^2(q^2-M^2_Z)^2}\right]      \eq

\newpage

\bqa  A_{FB,\mu}(q^2)&=&A^{Born}_{FB,\mu}(q^2)\bigm\{1+\nonumber\\
&&{q^4-\kappa^2(q^2-M^2_Z)^2
\over\kappa^2(q^2-M^2_Z)^2+q^4}\left[
\tilde{\Delta}_{\alpha}(q^2)+R(q^2)\right]
+{q^4\over\kappa^2(q^2-M^2_Z)^2+q^4}V(q^2)\bigm\} \eqa
\noindent
where
\bq
A^{Born}_{FB,\mu}(q^2)= {3q^2\kappa(q^2-M^2_Z)
\over2[ q^4+\kappa^2(q^2-M^2_Z)^2]}      \eq
\noindent

\bqa  &&\sigma_{5}(q^2)=\sigma^{Born}_{5}(q^2)\bigm\{1+
[{2(q^2-M^2_Z)^2\over
0.81q^4+0.06q^2(q^2-M^2_Z)+(q^2-M^2_Z)^2}][\tilde{\Delta}
_{\alpha}(q^2)]\nonumber\\
&&-[{0.81q^4\over 0.81q^4+0.06q^2(q^2-M^2_Z)+(q^2-M^2_Z)^2}]
[2R(q^2)+1.1V(q^2)]\nonumber\\
&&+[{0.06q^2(q^2-M^2_Z)\over 0.81q^4+0.06q^2(q^2-M^2_Z)+(q^2-M^2_Z)^2}]
[\tilde{\Delta}_{\alpha}(q^2)-R(q^2)-24.39V(q^2)]\bigm\}  \eqa

\noindent
where

\bqa 
\sigma^{Born}_{5}(q^2)
&\simeq&{44\pi\alpha^2\over9q^2}\left[1+0.81{q^4\over(q^2-M^2_Z)^2}
+0.06{q^2\over q^2-M^2_Z}\right]
\eqa

and

\bqa
\tilde{\Delta}^{(AGC)}_{\alpha} (q^{2}) =
-8 \pi \alpha \, \frac{q^{2}}{\Lambda^{2}} \,
\left[ f^{r}_{DW} + f^{r}_{DB} \right]
\eqa

\bqa
R^{(AGC)}(q^{2}) = 8 \pi \alpha \, \frac{(q^{2}-M^{2}_{z})}{\Lambda^{2}}
\,
\left[ \frac{1-\sweff}{\sweff} \, f^{r}_{DW} +
\frac{\sweff}{1-\sweff} \, f^{r}_{DB} \right ]
\label{XL}
\eqa

\bqa
V^{(AGC)}(q^{2}) = 8 \pi \alpha \, \frac{(q^{2}-M^{2}_{z})}{\Lambda^{2}}
\,
\left[ \frac{\sqrt{1-\sweff}}{\sweffu} \, f^{r}_{DW} -
\frac{\sweffu}{\sqrt{1-\sweff}} \, f^{r}_{DB} \right ]
\label{XLI}
\eqa



With Eqs.(6)-(9) and (16)-(18) at our disposal, we have moved to the 
practical task of deriving bounds for the four involved parameters in 
the (conventional) 
hypothesis that the new physics scale $\Lambda$ is 1~TeV, by 
following the same attitude adopted in Ref.\cite{gm2}. 
With this purpose, we have divided our input data into two sets. 
The first one consists of the two LEP1 and of the APV measurements. 
For these, we have used the experimental results quoted in 
Refs.\cite{LEP1,APV} with the related error and the SM predictions 
corresponding to $m_t=173.8$~GeV (according to the latest combined CDF/D0 
result \cite{latestmt}), $m_H=300$~GeV, 
$\alpha^{-1}_{QED}(M_z^2)=128.923$ \cite{DH} and $\alpha_s(M_z^2)=0.118$
(by following the electroweak working group choice \cite{LEP1}). 
To be more precise, the inputs of our analysis are:

\bq
\begin{array}{llllllll}
\Gamma^{exp}_l & = & 83.91  & \pm & 0.10~{\rm MeV} & & &  (~83.91~\rm MeV~)\\
A^{exp}_{FB,l} & = & 0.0171 & \pm & 0.0010         & & &  (~0.0151~)\\
Q^{exp}_W      & = & -72.11 & \pm & 0.27           & & & (~73.11 \pm 0.89~)
\end{array}
\eq
where the values in brackets are the SM predictions. The second set of
experiments consists of the $W$ mass determination and of the measurements
of the three LEP2 observables ($\sigma_{\mu}$, $A_{FB,\mu}$ and $\sigma_5$).
Here we have assumed that the final experimental values will agree with 
the SM predictions; to compute the latter quantities we have used the 
semianalytical program PALM, that was illustrated in a previous
paper~\cite{Bosonic}, where the ``Z-peak subtracted'' approach is
systematically adopted. Concerning the errors, we 
have used as final precision on the $W$ mass 30 MeV and we have defined 
the experimental uncertainties on the measurements at LEP2  
as the statistical errors achieved with an overall integrated luminosity 
of about $500~pb^{-1}$ collected by four experiments. 
Since the sensitivity of $\sigma_{\mu}$, $A_{FB,\mu}$ and $\sigma_5$
depends on the centre of mass energy, $E_{cm}$, we worked in the 
realistic scenario of several 
measurements performed in the energy range between 130~GeV and 200~GeV 
with a statistical significance determined, for each data sample, according
to the on going LEP operation and to the possible 
developments in the next two years of run. Namely, we assumed the following
centre of mass energy scan:
\bq
\begin{array}{lllllllll}
E_{cm}&= & 133 & 161 & 172 & 183 & 190 & 200 & ({\rm GeV})\\
{\cal L}_{int} &= & 10 &  10 &  10 &  50 & 200 & 250 & ({\rm pb^{-1}})
\end{array}
\eq
Technically, the results of our analysis have been obtained by minimising, 
in a conventional minimisation program, the $\chi^2$ variable
\bq
\chi^2 = \sum_{j=1}^4 \left(\frac{{\cal O}_j^{th} -{\cal O}_j^{exp}}
                           {\delta{\cal O}_j}\right)^2 +
   \sum_{j=1}^3 \sum_{k=1}^6 \left(\frac{{\cal O}_{jk}^{th} -
                                  {\cal O}_{jk}^{exp}}
                           {\delta{\cal O}_{jk}}\right)^2 
\eq
where the index $j$ runs over the seven observables, while the index $k$ 
runs over the six samples of data collected at LEP2. 
In each term of the $\chi^2$, the theoretical expression ${\cal O}^{th}$ 
consists of the sum of the SM prediction and of the shift induced by the 
AGC parameters. 

For the seek of our study, which aims to estimate the ultimate constraints
achievable in the case of negative experiment on the assumed set of
non-blind AGC, what is relevant is both the experimental accuracy 
of each measurement and the inherent sensitivity of the observables 
to every anomalous coupling. 
These two ingredients of the analysis affect
the shape of the $\chi^2$ around the minimum but not the location of the 
minimum itself. On the other hand, the most probable 
values of the AGC parameters, which can be determined only when the 
final experimental results will be available, will depend  
on $m_t$ and $m_H$, as will be briefly discussed later. In the following we
assume the latest combined CDF/D0 measurement of the top mass ($173.8\pm
3.2\pm 3.9$ \cite{latestmt}) and $m_H=$300~GeV.
Although one expects that the \underline{bounds} that we derive 
should be rather stable (unless unexpected strong variations from the
predicted accuracies will occur), clearly, the experimental inputs can 
(and will) be easily modified as soon as the real final data will be 
announced. This would allow to exclude (or detect) anomalies in the gauge
self-interaction sector by properly taking into account also the
uncertainty arising from $m_t$ 
and $m_H$, which hopefully will have reached a negligible level. 

The results of the overall analysis, made using the seven experimental
"data" and errors and minimizing the $\chi^2$ with respect to the four AGC
parameters at a time, are shown in Table \ref{tab1}. 

To be more precise, in the first row of Table \ref{tab1} 
we have listed the bounds that would be obtained by only using the four 
LEP1, APV, $M_W$ results. To shorten our notations, we shall call these 
data ``low energy'' data. In the second row, we give the results that 
would be derivable by adding to the previous ``low energy'' information 
that coming from the three LEP2 measurements. 
In Table \ref{tab2} and \ref{tab3} we report the error correlation 
matrices that correspond to the two cases.
Here we have defined $\delta f_{DW}, \delta f_{BW}, \delta f_{DB}$ and
$\delta f_{\phi,1}$ as the distance from the minimum of the
hyperplane corresponding to $\chi^2 = \chi^2_{min}+1$.

As one sees from inspection of Table \ref{tab1}, the addition of LEP2
data systematically ameliorates the general bounds. In particular, a
strong improvement is obtained for $f_{DB}$ and $f_{BW}$ (a factor
4-5). For the remaining parameters, a smaller but still remarkable (a
factor 2) reduction of the error bound is derived. 
To get a more specific feeling of the role of the different measurements,
we have first plotted in Fig. \ref{low-high} the contours, in the six
two-dimensional planes, corresponding to the bounds given in Table
\ref{tab1}, that is the projections in each plane of the $\chi^2 = 
\chi^2_{min}+1$ hyperplane. 
Here, as well as in all the plots presented, the artificial central 
values of the fits have been shifted to zero ``by hand'' in order to 
concentrate the attention on the significance of the result obtained 
with different sets of experimental inputs. 
The plot shows again, in a more immediate way, the relevance of the 
addition of the three LEP2 measurements on \underline{all} the four 
parameters, including the two ones that in our 
approach do not appear in the related theoretical expressions
(i.e. $f_{BW}$, $f_{\Phi, 1}$). This happens as a consequence of the
correlation among the parameters in the theoretical expressions of the
``low energy'' observables entering the $\chi^2$. 
In particular, from Table \ref{tab2}, one can conclude that 
$f_{DB}$, which is strongly correlated to the two energy insensitive 
parameters, drives the overall improvement.  

We considered also the occurrence that one, two or three AGC parameters 
are zero; the bounds achievable for the surviving parameters are listed in
Table \ref{tab<4fp}. Although there isn't any specific theoretical argument 
in favour of these scenarios, the results of this study clarify the 
interplay between the four parameters which results in the final 
correlations (Tables \ref{tab2} and \ref{tab3}). 

Finally, in Table \ref{68cl} we give the 68.3$\%$ C.L. bounds for the
four-free parameter fit. Since our $\chi^2$ is a quadratic function of the 
anomalous couplings, the shape of the region in the parameter space around
the minimum with a specific probability content does not depend on the
C.L. chosen. A comparison with Table \ref{tab1} shows that the 
region of the AGC parameter space allowed with a C.L. of 68.3\% corresponds
to a scaling by a factor 2.2 of the contours (Fig. \ref{low-high}) 
previously determined with our ``work definition'' of the bounds. 

The next question that we have addressed is that of understanding which ones of
the three LEP2 measurements that we have considered are more relevant. To
answer this point, we have plotted in Fig. \ref{noallhigh} the bounds that 
would be obtained by releasing \underline{one} out of the three LEP2 
measurements in the overall (``low-energy'' + LEP2) bound derivation. 
One sees in fact from those exclusion plots that the bulk of the 
information is provided by the addition to the low energy data of the two 
LEP2 cross sections; on the contrary, the
role of the muon asymmetry appears to be, in this specific context,
marginal, even at the most optimistic level of experimental accuracy.

Having stressed the relevance of the three LEP2 measurements for 
the derivation of meaningful bounds, we have then
studied the relative relevance of the four remaining
``low energy'' data. In other words, we have considered the four different
results that would be obtained by neglecting, each time, one of the four
low energy informations. The results are shown in Fig. \ref{noallow}, again
showing the projections on the six parameter planes.

As one sees from Fig. \ref{noallow}, the relevance of the three
measurements of $\Gamma_l$, $A_{FB, \mu}(M_Z^2)$ and $M_W$ is essentially
similar on all the six pairs: neglecting \underline{one} of these three
measurements introduces in the bounds different (appreciable)
comparable shifts. On the contrary, the role of the APV measurement seems
in this respect quite negligible, at the present level of overall
(experimental and theoretical) accuracy Eq.(21). In fact, a consistent 
reduction of the present uncertainty on $Q_W$ would be required for an 
effective impact of this observable in the analysis. Actually, a decrease 
of the error by a factor of 2 would still only marginally improve the bound 
on $f_{BW}$ ($7\%$), while a suppression by a factor of 5 is needed in
order to achieve a $33\%$ improvement of the bound on $f_{BW}$ and, as a
result of the correlation, a $28\%$ decrease of the allowed interval
for $f_{DB}$. 
 
A final comment concerns the effect of the top and Higgs masses
uncertainties in the result of our analysis. 
The values of $m_t$ and $m_H$ enter the theoretical expression of the 
two LEP1 observables and of $M_W$, and consequently affect the 
minimum of the $\chi^2$ and, therefore, the central values of the 
allowed intervals for the four anomalous couplings. 
At the present level of experimental precision on $A_{FB,l}$ and
$\Gamma_l$ and of the foreseen final error on $M_W$, the shifts induced in
the corresponding SM predictions by the error on $m_t$ (5~GeV) 
and by the uncertainty on $m_H$ are sizable. For example, when 
$m_H=$300~GeV, moving the value of $m_t$ from the central measured
value by one sigma produces a shift in $\sweffl(SM)$ equal to 30\% of the
experimental error arising from the measurement of $A_{FB,l}$; the 
corresponding shift of $\Gamma_l(SM)$ amounts to 50\% of the experimental 
error and the forecast for $M_W(SM)$ is moved slightly more than 
30~MeV. 
On the other hand, at a fixed value of the top mass, as $m_H$ ranges between
90~GeV and 1~TeV, the prediction on both $\sweffl$ and $\Gamma_l$
varies by roughly 
twice the present experimental errors and the variation of the
$M_W$ value is about 150~MeV. Those uncertainties reflect on a not 
negligible shift of the $\chi^2$ minimum in the purely low energy 
analysis. 
Namely, we observe a linear drift of the central values of
$f_{DW},~f_{DB}$ and $f_{\Phi,1}$ with $m_t$. A one sigma shift of $m_t$ 
moves them respectively by 10\%, 2\% and 5\% of the corresponding 68\%
C.L. errors. One can observe, therefore, that the future precision
in the determination of $m_t$ at the run II of TEVATRON, which is planned
to be $\delta m_t < 2$~GeV per experiment \cite{vancouver}, will play a 
very 
significant role in the definite bounds we will be able to derive on the 
AGC parameters. 
Of course, far less predictable is the impact of the outcomes on the Higgs
mass of future experiments.
Nevertheless, it's worth to point out that  
in our formulation the LEP2 observables are essentially free of
$m_t$, $M_H$ whose dominant contributions are reabsorbed in the theoretical
input (as discussed in~\cite{Zsub} this does not introduce any
\underline{appreciable} theoretical error). 
Since the major contributions to the bounds comes from LEP2 data, although
they bring direct information only on $f_{DW}$ and $f_{DB}$, we expect that 
the role of $m_t$, $M_H$ will be strongly weakened in the final
analysis to be performed. 

In conclusion, the most accurate determination of the bounds on the four
``non-blind'' parameters Eq.(1) appears to be that derivable from an
analysis of LEP1, LEP2 data combined with the experimental value
of $M_W$. As soon as the final LEP2 results will be established, our
analysis will be straightforwardly adapted to provide the final central
values to be used in the conclusive formulation.

\newpage

\begin{table}[htb]
\begin{center}
\caption{Bounds on the anomalous gauge
couplings obtained with a combined fit of present and future experimental
data. The defintion of the parameter uncertainties adopted here is the 1
$\sigma$ error in the $\chi^2$ minimization. 
{\bf Low} refers to the results from LEP1, APV and from the measurement of 
$M_W$, {\bf High} to the cross-section and asymmetry measurements at LEP2.
\label{tab1}}
\vskip 1cm
\begin{tabular}{c|cccc}
     & $\delta f_{DW}$  & $\delta f_{BW}$       & $\delta f_{DB}$       &
$\delta f_{\Phi, 1}$ \\ 
\hline \\
{\bf Low}  & 0.28 & 1.43 & 6.27 & 0.088 \\
{\bf Low+High} & 0.18 & 0.32 & 1.15 & 0.035 
\end{tabular}
\end{center}
\end{table}
\newpage

\begin{table}[htb]
\begin{center}
\caption{Correlation matrix from low energy data.\label{tab2}}
\vskip 0.3cm
\begin{tabular}{c|cccc}
     & $f_{DW}$ & $f_{BW}$      & $f_{DB}$      & $f_{\Phi, 1}$ \\      
\hline \\
$f_{DW}$        & 1 & 0.000 & -0.204 & -0.085 \\
$f_{BW}$      &   & 1     & -0.960 & 0.975 \\
$f_{DB}$      &   &       & 1      & -0.883 \\
$f_{\Phi, 1}$   & & & & 1
\end{tabular}
\end{center}
\end{table}
\newpage
\begin{table}[htb]
\begin{center}
\caption{Correlation matrix from low+high energy data.\label{tab3}}
\vskip 0.3cm
\begin{tabular}{c|cccc}
     & $f_{DW}$ & $f_{BW}$      & $f_{DB}$      & $f_{\Phi, 1}$ \\      
\hline \\
$f_{DW}$        & 1 & 0.049 & -0.785 & -0.114 \\
$f_{BW}$        &   & 1     & -0.351 & 0.930 \\
$f_{DB}$        &   &       & 1      & -0.077 \\
$f_{\Phi, 1}$   & & & & 1
\end{tabular}
\end{center}
\end{table}

\newpage

\begin{table}[htb]
\begin{center}
\caption{Bounds on the AGC parameters achieved in case of a reduced number
  of free parameters. \label{tab<4fp}}
\vskip 0.5cm
\footnotesize{
\begin{tabular}{|c|cc|}
\hline
1 free p. &         $f_{BW}=f_{DB}=f_{\Phi,1}=0$ &
                    $f_{DW}=f_{DB}=f_{\Phi,1}=0$ \\
(low) & $\delta f_{DW} = 0.14$ & 
                           $\delta f_{BW} = 0.067$ \\
(low+high) & $\delta f_{DW} = 0.094$ & 
                           $\delta f_{BW} = 0.067$ \\
\hline
1 free p. &         $f_{DW}=f_{BW}=f_{\Phi,1}=0$ &
                    $f_{DW}=f_{BW}=f_{DB}=0$ \\
(low) & $\delta f_{DB} = 0.53$ & 
                           $\delta f_{\Phi,1} = 0.008$ \\ 
(low+high) & $\delta f_{DB} = 0.42$ & 
                           $\delta f_{\Phi,1} = 0.008$ \\ \hline
\end{tabular}
}
\vspace{0.5cm}
\footnotesize{
\begin{tabular}{|c|ccc|}
\hline
 2 free p. & $f_{DB}=f_{\Phi,1}=0$ & 
             $f_{BW}=f_{\Phi,1}=0$ &
             $f_{BW}=f_{DB}=0$ \\
(low) & $\delta f_{DW}=0.22\:\:\delta f_{BW}=0.10$ & 
        $\delta f_{DW}=0.25\:\:\delta f_{DB}=0.94$ &
        $\delta f_{DW}=0.19\:\:\delta f_{\Phi,1}=0.010$ \\
(low+high) & $\delta f_{DW}=0.11\:\:\delta f_{BW}=0.077$ &
         $\delta f_{DW}=0.16\:\:\delta f_{DB}=0.74$ & 
         $\delta f_{DW}=0.10\:\:\delta f_{\Phi,1}=0.089$ \\ \hline
 2 free p. & $f_{DW}=f_{\Phi,1}=0$ & 
             $f_{DW}=f_{DB}=0$     & 
             $f_{DW}=f_{BW}=0$    \\
(low) & $\delta f_{BW}=0.30\:\:\delta f_{DB}=2.36$ &
        $\delta f_{BW}=0.28\:\:\delta f_{\Phi,1}=0.034$ &
        $\delta f_{DB}=1.20\:\:\delta f_{\Phi,1}=0.018$ \\
(low+high) & $\delta f_{BW}=0.11\:\:\delta f_{DB}=0.69$ &
             $\delta f_{BW}=0.28\:\:\delta f_{\Phi,1}=0.034$ &
             $\delta f_{DB}=0.62\:\:\delta f_{\Phi,1}=0.012$ \\ \hline
\end{tabular}
}

\vspace{0.5cm}
\footnotesize{
\begin{tabular}{|c|cc|}
\hline
 3 free p.       & $f_{DW} = 0$ & 
                           $f_{BW} = 0$ \\ 
(low) & $\delta f_{BW}=1.44\:\delta f_{DB}=6.14\:\delta f_{\Phi,1}=0.087$ &
        $\delta f_{DW}=0.28\:\delta f_{DB}=1.75\:\delta f_{\Phi,1}=0.020$ \\
(low+high) 
      & $\delta f_{BW}=0.32\:\delta f_{DB}=0.71\:\delta f_{\Phi,1}=0.035$ &
        $\delta f_{DW}=0.18\:\delta f_{DB}=1.08\:\delta f_{\Phi,1}=0.013$
 \\ \hline
 3 free p.       & $f_{DB} = 0$ & 
                           $f_{\Phi,1} = 0$ \\ 
(low) & $\delta f_{DW}=0.27\:\delta f_{BW}=0.40\:\delta f_{\Phi,1}=0.041$ &
        $\delta f_{DW}=0.27\:\delta f_{BW}=0.32\:\delta f_{DB}=2.95$ \\ 
(low+high) 
      & $\delta f_{DW}=0.11\:\delta f_{BW}=0.30\:\delta f_{\Phi,1}=0.035$ &
        $\delta f_{DW}=0.18\:\delta f_{BW}=0.12\:\delta f_{DB}=1.15$ \\ 
\hline
\end{tabular}
}
\end{center}
\end{table}
\newpage

\begin{table}[htb]
\begin{center}
\caption{68.3\% Confidence Level bounds on the anomalous gauge
couplings. As before, {\bf Low} refers to the LEP1, APV and $M_W$ data, 
{\bf High} to the LEP2 measurements.\label{68cl}}
\vskip 1cm
\begin{tabular}{c|cccc}
     & $\delta f_{DW}$  & $\delta f_{BW}$       & $\delta f_{DB}$       &
$\delta f_{\Phi, 1}$ \\ 
\hline \\
{\bf Low}  & 0.58 & 3.00 & 13.12 & 0.184 \\
{\bf Low+High} & 0.38 & 0.68 & 2.42 & 0.074 
\end{tabular}
\end{center}
\end{table}

\newpage

\begin{center}

{\large \bf Figure captions}
\end{center}
\vspace{0.5cm}

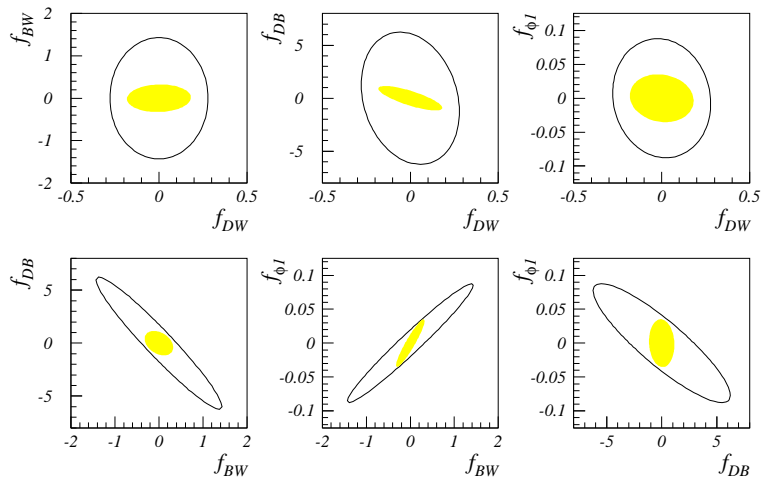
{\bf Fig.1:} 
Projection of the $\chi^2<\chi^2_{min}+1$ region in the 
space of the four anomalous gauge couplings onto the six possible 
coordinate planes. The outer ellipses are obtained from low energy data only;
the inner ones, by including the LEP2 data.

\vskip 1cm

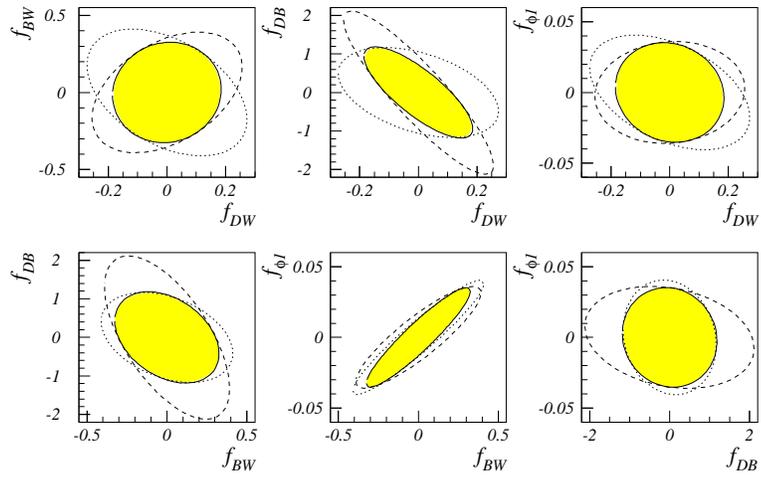
{\bf Fig.2:} 
Projected ellipses obtained by a global fit to the full set of
measurements (LEP1, LEP2, APV and $M_W$) releasing one of the {\bf high}
energy constraints. The three curves correspond respectively to the 
exclusion of $\sigma_\mu$ (dashed), 
$\sigma_5$ (dotted), $A_{FB, \mu}$ (solid). 
The most internal region (shaded) is the overall result.

\vskip 1cm

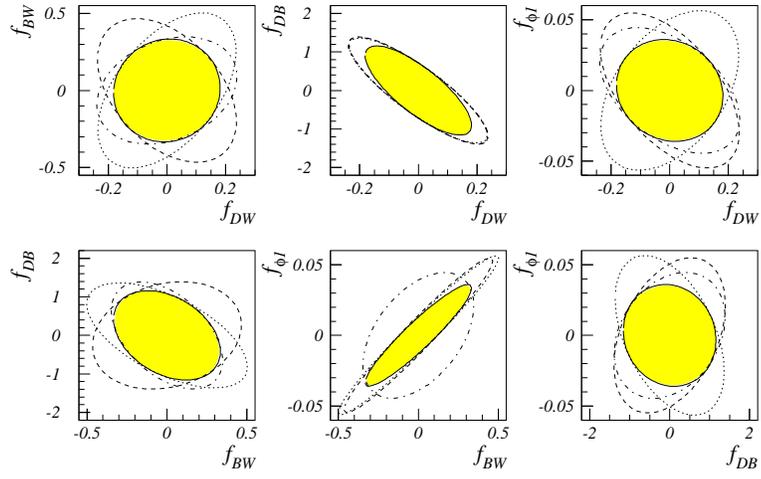
{\bf Fig.3:} 
Projected ellipses obtained by a global fit to the full set of
measurements (LEP1, LEP2, APV and $M_W$) releasing one of the {\bf low} 
energy constraints. 
The four curves correspond respectively to the exclusion of
$A_{FB, l}$ (dashed), $\Gamma_l$ (dotted), $Q_W$ (solid) and
$M_W$ (dot-dash). The most internal region (shaded) is the overall
result.

\newpage
\begin{figure}[htb]
\vspace{15cm}
\includegraphics{ps1}
\end{figure}

\newpage

\begin{figure}[htb]
\vspace{9cm}
\includegraphics{ps2}
\end{figure}

\newpage

\begin{figure}[htb]
\vspace{9cm}
\includegraphics{ps3}
\end{figure}

\end{document}